 \documentclass[12pt,preprint]{aastex}

\newcommand{\HESS}{H.E.S.S.}
\newcommand{\this}{HESS\,J1640--465}
\newcommand{\asca}{AX\,J1640.7--4632}
\newcommand{\xmm}{XMMU\,J164045.4--463131}
\newcommand{\eg}{3EG\,J1639--4702}
\newcommand{\snr}{G338.3--0.0}

\slugcomment{Submitted to the Astrophysical Journal}

\shorttitle{{\emph{XMM-Newton}} observations of \this}
  \shortauthors{S.~Funk et al.}

\begin{document}


\title{{\emph{XMM-Newton}} observations reveal the X-ray counterpart
of the very-high-energy $\gamma$-ray source \this}


\author{S.~Funk\altaffilmark{1,2},
  J.~A.~Hinton\altaffilmark{3},
  G.~P\"uhlhofer\altaffilmark{4}, 
  F.~A.~Aharonian\altaffilmark{5,2},
  W.~Hofmann\altaffilmark{2},
  O.~Reimer\altaffilmark{6},
  S.~Wagner\altaffilmark{4},
}
\altaffiltext{1}{Kavli Institute for Astroparticle Physics and
  Cosmology, SLAC, CA-94025, USA}
\altaffiltext{2}{Max-Planck-Institut f\"ur Kernphysik, P.O. Box
  103980, 69029 Heidelberg, Germany}
\altaffiltext{3}{School of Physics \& Astronomy, University of Leeds,
  Leeds LS2 9JT, UK} 
\altaffiltext{4}{Landessternwarte, Universit\"at Heidelberg,
  K\"onigstuhl, 69117 Heidelberg, Germany}
\altaffiltext{5} {Dublin Institute for Advanced Studies, 5 Merrion
  Square, Dublin 2, Ireland} 
\altaffiltext{6}{Stanford University, HEPL \& KIPAC, Stanford, CA
  94305-4085, USA}

\begin{abstract}
  We present X-ray observations of the as of yet unidentified very
  high-energy (VHE) $\gamma$-ray source \object{\this}\ with the aim
  of establishing a counterpart of this source in the keV energy
  range, and identifying the mechanism responsible for the VHE
  emission.  The 21.8~ksec XMM-Newton observation of \this\ in
  September 2005 represents a significant improvement in sensitivity
  and angular resolution over previous ASCA studies in this region.
  These new data show a hard-spectrum X-ray emitting object at the
  centroid of the H.E.S.S. source, within the shell of the radio
  Supernova Remnant (SNR) \snr.  This object is consistent with the
  position and flux previously measured by both ASCA and Swift-XRT but
  is now shown to be significantly extended. We argue that this object
  is very likely the counterpart to \this\ and that both objects may
  represent the Pulsar Wind Nebula of an as of yet undiscovered pulsar
  associated with \snr.
\end{abstract}


\keywords{ISM: Supernova remnants, plerions -- ISM: individual
  objects: \object{\this}, \object{\snr},
  \object{\asca}, \object{\eg} --
  X-rays: observations -- gamma-rays: observations}

\section{Introduction}
The last years brought an emerging new population of Galactic VHE
$\gamma$-ray sources. More than 20 new objects are now established as
sources of $\gamma$-rays above 100~GeV by Atmospheric Cherenkov
Telescopes such as \HESS~\citep{Hinton}. In particular a first
unbiased survey of the Inner Galaxy by \HESS\ found more than a dozen
new sources~\citep{HESSScan, HESSScanII}. While several Galactic
$\gamma$-ray sources can be identified with counterparts at other
wavebands, such as \object{RX\,J1713.7--3946}~\citep{HESSRXJ1713,
HESSRXJ1713II} or \object{MSH--15-5\emph{2}}~\citep{HESSMSH}, most of
them have not yet been identified with a striking positional
counterpart~\citep{HESSScanII, FunkBarcelona}. A counterpart search
proceeds in several steps: first, a positional counterpart at other
wavebands must be established, typically through high-angular
resolution observations in the radio and X-ray bands. In these bands
non-thermal particle populations can be studied through their
synchrotron emission without being strongly absorbed by interstellar
dust. Following the detection of an astrometric counterpart, a firm
identification requires both a viable $\gamma$-ray emission mechanism
and a consistent multi-frequency picture of the positional
counterpart. Here we report on X-ray observations on \this\ with the
X-ray {\emph{XMM-Newton}} satellite as part of an ongoing programme to
study unidentified VHE $\gamma$-ray sources with X-ray satellites.

\this\ was first detected by \HESS\ in a survey of the inner
Galaxy~\citep{HESSScan, HESSScanII} in 2004. Following the initial
discovery the source was re-observed in pointed observations with a
resulting statistical significance of $\sim 14 \sigma$ in 14.3 hours
of lifetime, with a total of 313 $\gamma$-ray events above the
background of $\sim 300$ events. The $\gamma$-ray emission profile is
Gaussian and rather compact in nature with an rms width of $2.7\arcmin
\pm 0.5\arcmin$, less extended than most of the sources detected in
the survey. The $\gamma$-ray energy spectrum was reported to follow a
simple power-law with photon index $2.42 \pm 0.15$ and a total
integrated flux above 200~GeV of $(20.9 \pm 2.2_{\mathrm{stat}})
\times 10^{-12}$ cm$^{-2}$ s$^{-1}$ (corresponding to $2.2 \times
10^{-11}$ erg cm$^{-2}$ s$^{-1}$). \this\ is located in the Galactic
plane at an RA-Dec$_{2000}$ of
$16^{\mathrm{h}}40^{\mathrm{m}}44^{\mathrm{s}}$,
$-46^{\mathrm{d}}31\arcmin57\arcsec$ (or $l=338.32^{\circ},
b=-0.02^{\circ}$). This direction on the sky lines up with the
prominent 3-kpc-arm-tangent region of the Galaxy at a distance of
8~kpc.

\this\ was found to show compelling positional coincidence with the
shell-type Supernova remnant (SNR) \snr\ \citep{Green}. This
broken-shell SNR with a diameter of 8\arcmin\ was detected in the
843~MHz radio survey using the Molonglo Observatory Synthesis
Telescope (MOST)~\citep{Molonglo}. \snr\ is located at a
RA-Dec$_{2000}$ position of
$16^{\mathrm{h}}41^{\mathrm{m}}00^{\mathrm{s}}$,
$-46^{\mathrm{d}}34\arcmin$ and is not particularly well studied by
radio telescopes~\citep{Green}. A reanalysis of the MOST data
~(T.~Cheung, private communication) yields a total radio emission from
the remnant of $\sim 10$~Jy on a background of $\sim 10$~mJy per beam
outside of the remnant. The central point source within the remnant
shows a flux of $\sim 160$~mJy with a peak flux of $\sim 100$~mJy. The
connection between this central source and the remnant is as of yet
not evident. The radio emission from the interior of the remnant away
from the central point-source amounts to $\sim 20$~mJy. No pulsar is
known within \snr\ in the current ATNF pulsar
catalogue~\citep{ATNF}. The projection of \snr\ is located at the edge
of a bright HII-region~\citep{Whiteoak} (located at a distance of
3~kpc) with ongoing star formation and dust emission. This region
seemingly connects the radio emission of \snr\ with the close-by
Green's catalogue SNR G\,338.5+0.1~\citep{Green}. The best distance
estimate of \snr\ so far comes from the relation of surface brightness
to distance which places \snr\ at a distance of 8.6~kpc, right within
the 3-kpc-arm-tangent region (and leaves it thus unrelated to the
HII-region). At X-ray energies the region has been observed in the
ASCA Galactic plane survey, resulting in the detection of \asca\ in
positional coincidence with the SNR~\citep{ASCA}. The ASCA source
showed a rather soft photon index of $3.0^{+1.1}_{-0.9}$, a total
unabsorbed flux between 2~keV and 10~keV of $F_{2-10\mathrm{~keV}} =
1.2\times 10^{-12}$ erg cm$^{-2}$ s$^{-1}$, and a rather high
absorption of n$_{\mathrm{H}} = 9.6^{+4.7}_{-3.3} \times 10^{22}
\mathrm{cm}^{-2}$ compared to the total column density in this
direction (n$_{\mathrm{Gal H}} = 2.2 \times 10^{22} \mathrm{cm}^{-2}$
as reported by ~\citet{DickeyLockman}). Recent {\emph{Swift}} XRT
observations on \this\ revealed one X-ray point-source within the
radio shell and in coincidence with \asca\ ~\citep[source
\#1 in][]{Swift1640}. The XRT data confirm the spectral parameters found
by ASCA resulting in still rather unconstrained values of the photon
index $\Gamma = 2.6^{+1.0}_{-1.1}$ and the 2--10~keV flux
$F_{2-10\mathrm{keV}} = 0.7\times 10^{-12}$ erg cm$^{-2}$ s$^{-1}$.

At $\gamma$-ray energies above 100~MeV, \eg, an unidentified EGRET
source~\citep{EGRETCat}, is found spatially compatible with \this\ at
an angular distance of 34\arcmin. \eg\ was only accepted for inclusion
in the 3EG catalogue based on its cumulative significance over 4 years
of EGRET observation, but never satisfied the detection threshold in
any individual EGRET observation. Consequently, there is no reliable
variability estimate available for this source, and a classification
based on the available EGRET observables appeared impossible.  \eg\
was labelled as an extended and confused $\gamma$-ray source, which
puts the rather tight positional coincidence of 34\arcmin\ with \this\
somewhat into perspective. Various positional counterparts were found:
a rather speculative suggestion was to associate it both with the
unidentified ASCA source AX\,J1639.0--4642~\citep{ASCA}, and the
INTEGRAL source IGR\,J16393--4643~\citep{MaliziaIntegral}, concluding
that \eg\ is a dust-enshrouded microquasar
candidate~\citep{Combi}. This suggestion was revised by the detection
of a heavily obscured X-ray pulsar in {\emph{XMM-Newton}} and
{\emph{RXTE}} observations coincident with
IGR\,J16393--4643~\citep{Bodaghee, Thompson}. The corresponding mass
estimates allowed for a classification of IGR\,J16393--4643 as a
high-mass X-ray binary system. \asca\ and IGR\,J16393--4643 might
therefore not be related to \eg\ given that this would constitute the
first association between an unidentified EGRET source and a highly
obscured INTEGRAL-detected HMXB. This scenario currently lacks both
viable multi-waveband modelling and predictions for GeV
energies. Other associations for \eg\ were suggested, including
positionally coincident radio pulsars (PSR\,J1637--4642,
PSR\,J1640--4648, and PSR\,J1637--4721) ~\citep{TorresButtCamilo} and
SNRs (G337.8--0.1, G338.1+0.4, and \snr)~\citep{Romero}. Two of these
pulsars can be disregarded on energetics grounds~\citep{TorresSNR},
leaving only PSR\,J1637--4721 with a moderate required efficiency of
12\% for conversion of spin down power into EGRET
$\gamma$-radiation. Finally, \citet{Romero} also listed the stellar
cluster and OB-associations NGC\,6204, and Ara\,1A~A as positional
coincidences, the latter additionally coinciding with 3EG\,J1655-4554.
Summarising the possible associations of \eg, the typical situation of
an observationally weakly constrained EGRET detection allows for a
variety of interpretations. A unique counterpart association will only
be possible if the source location uncertainty is refined (using
GLAST-LAT), or intriguing features at other wavebands are discovered,
as perhaps in this case from \HESS\ observations.

The positional coincidences of \this\ with \snr\ and \asca\ (and
possibly also \eg) suggests a connection between these
objects. However, from the described multi-frequency data it remains
so far unclear whether the X-ray and $\gamma$-ray emission originates
within the shell of the radio SNR or rather in connection with a
central X-ray object embedded within the shell. If the distance
estimate of 8.6~kpc is correct, the radius of the SNR is $\sim 10$~pc
and the Sedov solution for the evolutionary stage of the SNR yields
ages between $\sim (0.2 - 2 ) \times 10^{4}$ years (assuming a typical
ambient density between $\sim 0.1 - 10 $ cm$^{-3}$ and an explosion
energy of $E= 10^{51}$ erg). In particular for densities larger than
the canonical $1 \mathrm{cm}^{-3}$, the Sedov solution yields an age
in excess of the typical VHE-bright $\gamma$-ray SNRs, such as
RX\,J1713.7--3946 and RX\,J0852.0--4622 with ages of $\sim 2000$ years
or younger. In this letter we report on X-ray observations towards the
\snr/\this\ region performed with {\emph{XMM-Newton}} with the aim to
study the connection between the radio/X-ray and $\gamma$-ray emitting
regions in the light of higher angular resolution X-ray data.

\section{{\emph{XMM-Newton}} Observations of the region}
\begin{figure}
  \centering
  \includegraphics[width=0.65\textwidth]{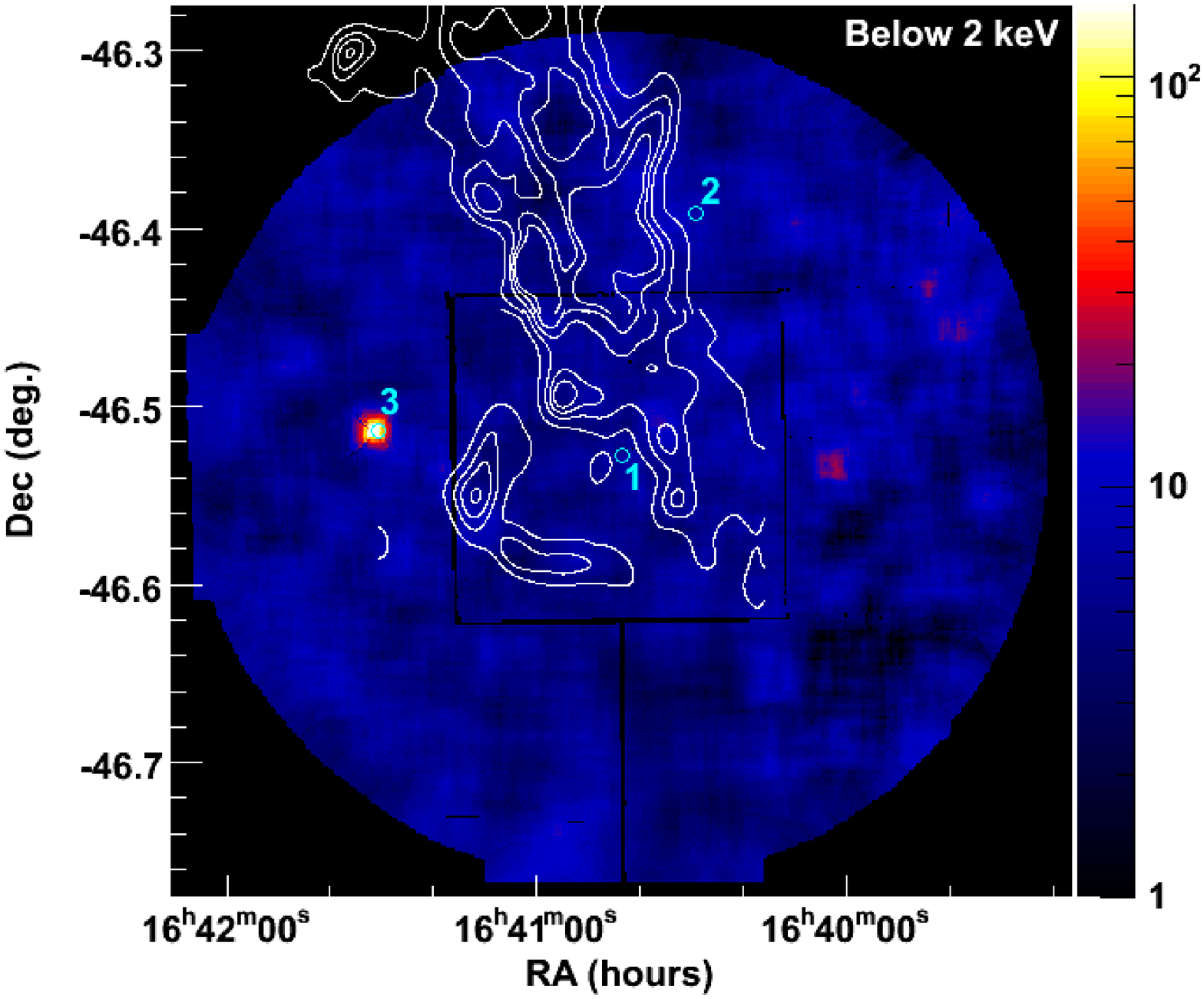}
  \includegraphics[width=0.65\textwidth]{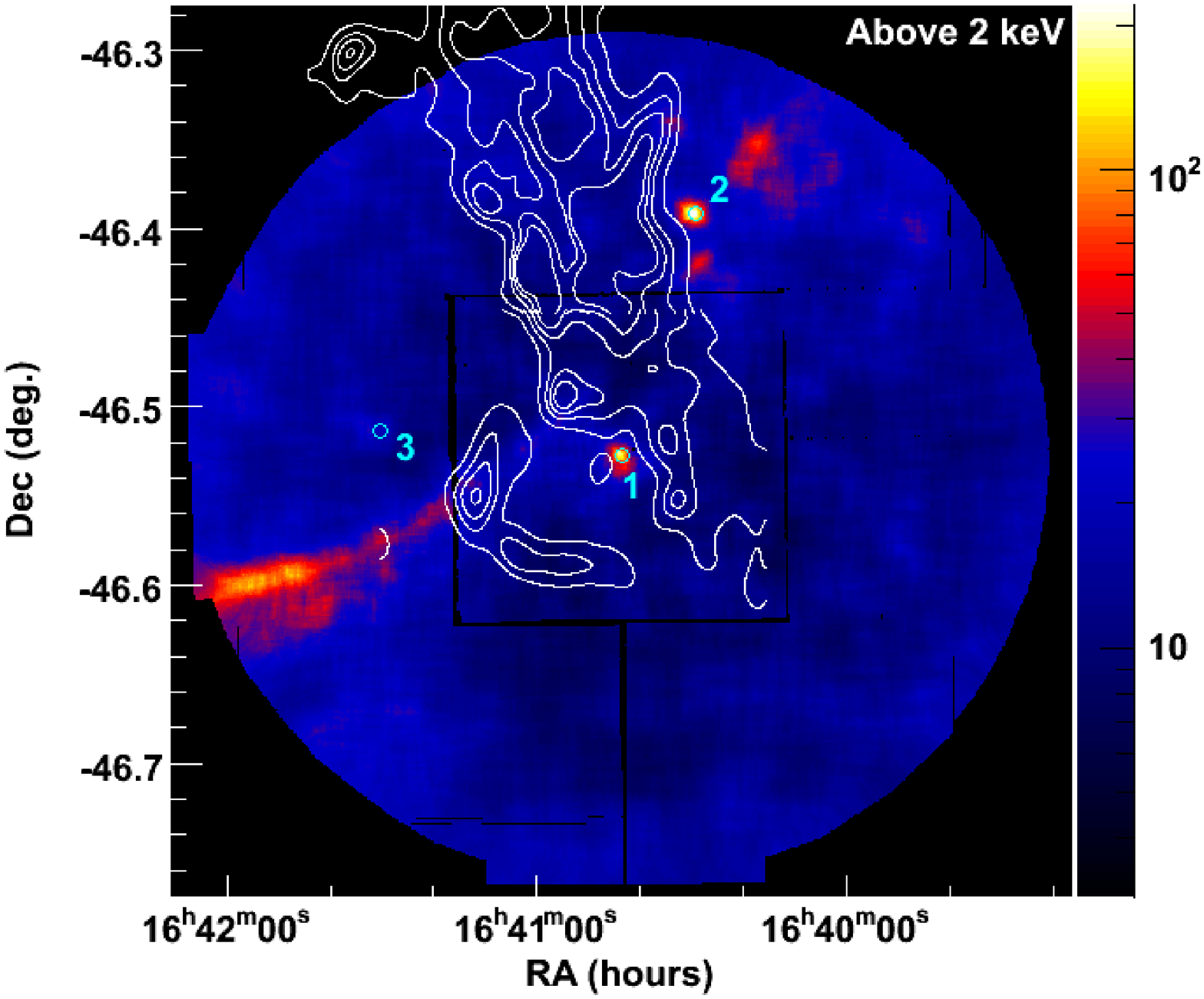}
  \caption{The {\emph{XMM-Newton}} field of view for the observation
    on \this. This composite image of the MOS-1 and MOS-2 count maps
    shows the energy range below 2~keV (top) and above 2~keV (bottom),
    the counts were adaptively smoothed. Three prominent X-ray sources
    can be seen lining up with previously reported ASCA and
    {\emph{Swift}} XRT sources (the cyan circles denote the position
    of the {\emph{Swift}} sources \#1, \#2, and \#3 as reported
    in~\citet{Swift1640}). Source \#1 is coincident with the known
    X-ray source \asca\ located within the radio shell of \snr. The
    843~MHz MOST radio data are shown as white contours. The weak
    central radio object within the shell is also apparent in these
    contours. Also visible from this image is stray-light
    contamination from the close-by low-mass X-ray binary GX\,340+0.
    \label{fig::XMMFoV}}          
\end{figure}

{\emph{XMM-Newton}} observed \this\ on the 20th of August 2005 for
21.8~ksec in satellite revolution 1043 (observation id
0302560201). All the instrument cameras (EPIC MOS1, MOS2, PN) were
operated in full-frame mode with a medium filter to screen out optical
and UV light. For data analysis, the {\emph{XMM-Newton}} Science
analysis software (SAS) version 7.0 was used together with the
Extended Source Analysis Software package (XMM-ESAS) version
1.0~\citep{Snowden}. Following standard data reduction and calibration
procedures, the data set was cleaned from temporally occurring
background caused by soft proton flares. The resulting observation
time amounts to only 7.3~ksec of useful data. Figure~\ref{fig::XMMFoV}
shows an adaptively smoothed count map below 2~keV (top) and above
2~keV (bottom) combining events for the MOS-1 and MOS-2 detectors of
{\emph{XMM-Newton}} of the region surrounding \this. The white
contours show the 843~MHz MOST-detected radio emission. Three
prominent X-ray sources are visible, coinciding with the recently
reported {\emph{Swift}} XRT sources (and accordingly labelled \#1,
\#2, and \#3). Source \#1 positionally coincides with the ASCA source
\asca\ and is located within the radio shell of \snr. No shell-like
X-ray emission such as seen in the MOST data is apparent in the
{\emph{XMM-Newton}} data set. Also visible is the band of stray-light
crossing the field of view from the north to the east, which can be
attributed to the close-by ($\sim 0.7^{\circ}$) low mass X-ray binary
GX\,340.0. This stray-light contribution to the spectral analysis
within a small region surrounding the X-ray source \asca\ seems to be
rather small, since various background estimation methods using
regions within the same field of view or from blank sky observations
yield results comparable within the statistical uncertainties.

\begin{figure}
  \centering
  \includegraphics[width=0.65\textwidth]{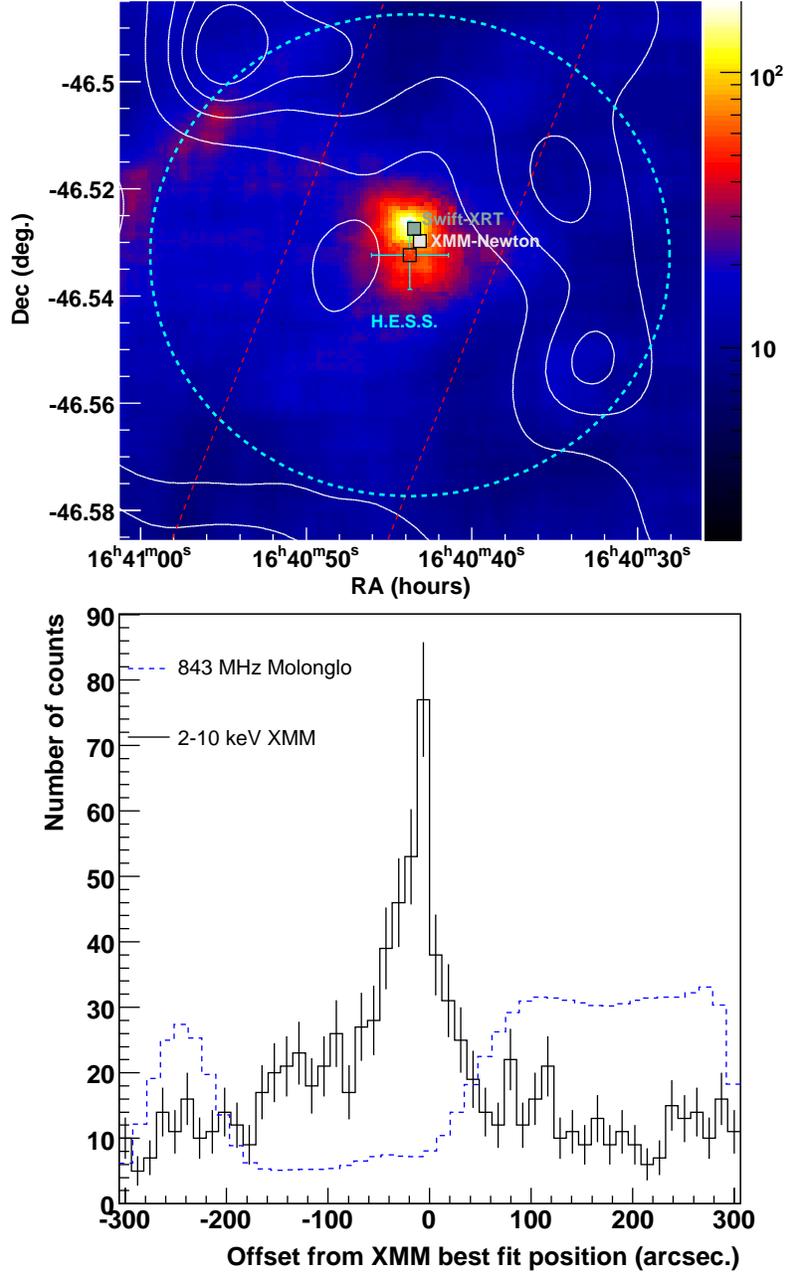}
  \caption{{\bf{Top:}} Zoomed view on the {\emph{XMM-Newton}} field of
    view for the observation on \this\ for events above 2~keV
    (composite image of the MOS-1 and MOS-2, adaptively smoothed). The
    white contours denote the 843~MHz MOST radio data showing the
    inner edge of the radio shell of \snr. The black square along with
    the solid cyan lines denote the best fit position (and $1 \sigma$
    error) of the VHE $\gamma$-ray source \this, the dashed cyan
    circle indicates the rms extension of \this. Also shown are the
    {\emph{XMM-Newton}} best fit position as detected with
    {\emph{emldetect}} as well as the best fit position of the
    {\emph{Swift}} XRT X-ray source.  {\bf{Bottom:}} Slice along the
    dotted red box (top Figure) through the X-ray (black) and radio
    (blue) emission. The difference between the shell-like structure
    in the 843~MHz radio data and the compact core with extended tail
    towards the south in the {\emph{XMM-Newton}} data is clearly
    apparent.
    \label{fig::XMMFoVZoom}}          
\end{figure}

In a first step to characterise the X-ray sources in the region, the
standard source detection algorithm {\emph{emldetect}} has been
applied to the data in the energy range from 0.5--10~keV, and as a
cross-check in the restricted energy bands 0.5--2~keV, 2--4.5~keV, and
4.5--10~keV. Three X-ray sources are detected as depicted in
Figure~\ref{fig::XMMFoV} by the cyan circles. The detection of source
\#1 in the 0.5--2~keV energy band is very weak, indicating a rather
hard source with a high degree of absorption at low energies. Source
\#3 on the other hand is only detected in the 0.5--2~keV band,
indicating a rather soft source. While these three sources are located
within the several-arcsecond error bars of the {\emph{Swift}}
XRT-sources~\citep{Swift1640}, the faint XRT-sources \#4 and \#5 are
not detected in this {\emph{XMM-Newton}} data, neither in the
MOS-detectors in which they fall into gaps between the CCD chips nor
in the PN-detector in which they are covered by the stray-light from
\object{GX\,340.0}. Table~\ref{tab::Properties} summarises the
properties of the three sources for the 0.5--10~keV energy band. Only
source \#1, coincident with the ASCA source \asca\ (in the following
named \xmm) is extended in nature, consistently in all energy
bands. Figure~\ref{fig::XMMFoVZoom} shows a zoom on the region
surrounding \xmm\ along with the best fit positions for the
{\emph{XMM-Newton}} data presented here and the {\emph{Swift}} XRT
data~\citep{Swift1640}. The source position fitting tool
{\emph{emldetect}} determines the extension of \xmm\ to be
incompatible with a point-source. Using a Gaussian model, the
extension in the best-constrained energy band above 4.5~keV is
determined to be $10.7\pm0.1$ image pixels, corresponding to a width
on the sky of $27\arcsec \pm 3\arcsec$. For the lower energy bands,
the best-fit width is larger but may be affected by stray-light, the
band 4.5~keV to 10~keV was therefore chosen to provide the most
reliable extension of the source. As is apparent in a slice through
the source, as shown in Figure~\ref{fig::XMMFoVZoom} (bottom), a
Gaussian profile might not be the correct representation of the source
morphology, which seems to exhibit a \emph{tail} extending mostly
southwards. As can also be seen from the slice, the contamination of a
possible point-source-like pulsar within the source is expected to
cover one or two bins in the slice and therefore to contaminate the
emission by $\sim 20\%$. The source position determined using the
{\emph{XMM-Newton}} data (see Table~\ref{tab::Properties}) is
compatible with both the ASCA source (at
$16^{\mathrm{h}}40^{\mathrm{s}}42.24^{\mathrm{s}}$,
$-46^{\circ}32\arcmin42.0\arcsec$ with a typical positional error of
1\arcmin\ on the ASCA position and a distance of 1.3\arcmin\ between
the ASCA and\ the {\emph{XMM-Newton}} best fit position) and the Swift
XRT source (at $16^{\mathrm{h}}40^{\mathrm{s}}43.5^{\mathrm{s}}$,
$-46^{\circ}31\arcmin38.6\arcsec$ with a positional error of 6\arcsec\
at a distance of 20\arcsec). As is apparent from
Figure~\ref{fig::XMMFoVZoom}, the extended source \xmm\ is
positionally coincident with \this\ and located within the radio shell
\snr, suggesting an association between the sources in these different
energy bands. \xmm\ does not appear to be positionally coincident with
the weak point-like radio source located within the shell of \snr\ if
the error on the radio position is indeed as small as several
arcseconds~\citep{Molonglo}. Future radio studies are well motivated
to identify the nature of this compact radio source.

\begin{table*}[h]
  \centering
    \caption{X-ray sources detected in observations of \this\ using
    the detection algorithm \emph{emldetect}. The parameters given
    here are for the energy range between 0.5~keV and 10~keV. The
    second column gives the name recommended by the
    {\emph{XMM-Newton}} SOC and the IAU for source detections. Columns
    3 and 4 give J2000 coordinates. Column 5 gives the error on the
    source position in arcseconds and column 6 gives the number of
    counts in EMOS1 and EMOS2 within a 35\arcsec\ integration region
    using events above 0.5~keV.\vspace{0.4cm} }
    \label{tab::Properties}
    \begin{tabular}{c | c | c c c | c }
      \tableline
      Id & XMMU\,J &  RA$_{2000}$ & Dec$_{2000}$ & Error RA$_{2000}$ & Counts \\
      & & (h:m:s) & (d:\arcmin:\arcsec) & (\arcsec) &  \\ \hline
      1 & 164045.4--463131 & 16:40:45.39 & --46:31:31.1 & 8.5 & 222\\
      2 & 164029.6--462328 & 16:40:29.58 & --46:23:28.0 & 1.5 & 153\\
      3 & 164131.0--463048 & 16:41:30.98 & --46:30:48.0 & 0.6 & 215\\ \tableline
    \end{tabular}
\end{table*}

A catalogue search did not yield any obvious counterparts to sources
\#2 and \#3. Since the angular distance to these two sources is too
large to be reasonably associated with the VHE $\gamma$-ray emission
in \this, the following analysis will be focused on \xmm. For the
spectral analysis, the data was extracted using SAS 7.0. The reduced
data set has been analysed and fitted using XSPEC (version
12.2.1). Several different background estimation techniques have been
applied, including determining the background from the same field of
view (both a ring around the source region and the source region
mirrored at the centre of the camera) as well as from blank field
observations. The different background estimation techniques agree
well within the statistical errors and therefore in the following the
background taken from a ring around the source in the same field of
view will be used. The application of a background estimate from the
same field of view has the advantage that all relevant backgrounds for
this observation are included, such as the X-ray emission from the
Galactic ridge, the residual particle background as well as the
instrumental background. To encompass the whole extension of the
source a source region of radius 75\arcsec\ has been used for the
spectral analysis. Consistent results with larger error bars were
achieved with source extraction regions of radius 50\arcsec\ and
100\arcsec. Table~\ref{tab::SpectralFitting} summarises the parameters
of the spectral fitting. Two spectral models were fitted to the data,
an absorbed power-law and an absorbed black-body spectrum.

\begin{table*}[h]
    \caption{{\emph{XMM-Newton}} spectral properties of \xmm\ for a
    simultaneous fit to the EMOS1, EMOS2, and EPN data. The two fit
    models are an absorbed power-law and an absorbed black-body. The
    fit parameters for the absorbed power-law are the absorption
    density n$_{\mathrm{H}}$ in units of cm$^{-2}$, the photon index
    $\Gamma$, and the normalisation at 1~keV (k$_\mathrm{1keV}$). The
    fit parameters for the black-body spectrum are the absorption
    density n$_{\mathrm{H}}$ again in units of cm$^{-2}$, the
    temperature $kT$ in units of keV, and the normalisation at 1~keV
    (k$_\mathrm{1keV}$).  Also given is the integrated absorbed
    (i.e. observed) flux between 2~keV and 10~keV
    F$_{\mathrm{2-10~keV}}$ (in units of erg $\mathrm{cm}^{-2}$
    s$^{-1}$ for the two models. The errors given correspond to 90\%
    confidence levels.\vspace{0.2cm}}
    \label{tab::SpectralFitting}
    \centering
    \begin{tabular}{ c | c c}
      \tableline
      Parameter & Value (powerlaw) & Value (black-body)\\ 
      \tableline
      EMOS1 Counts & \multicolumn{2}{c}{321} \\
      EMOS2 Counts & \multicolumn{2}{c}{290} \\
      EPN Counts & \multicolumn{2}{c}{807}\\
      n$_{\mathrm{H}}$ (cm$^{-2}$)& (6.1$^{+2.1}_{-0.6}) \times
      10^{22}$ & $(3.6^{+1.1}_{-0.8})\times 10^{22}$ \\ 
      $\Gamma$         & 1.74$^{+0.12}_{-0.10}$ & \\
      $kT$ (keV)            & & 1.56$^{+0.09}_{-0.19}$  \\
      k$_\mathrm{1keV}$ & $(2.6^{+1.4}_{-0.6}) \times 10^{-4}$ &
      $(1.1^{+0.08}_{-0.14}) \times 10^{-5}$ \\  
      F$_{\mathrm{2-10~keV}}$ (erg cm$^{-2}$ s$^{-1}$)& $6.6 \times
      10^{-13}$ & $6.4 \times 10^{-13}$\\ 
      $\chi^2$/d.o.f.  & 2702/3001 & 2694/3001\\
      \tableline
    \end{tabular}
\end{table*}

\begin{figure}
  \centering
  \includegraphics[width=0.6\textwidth, angle=270]{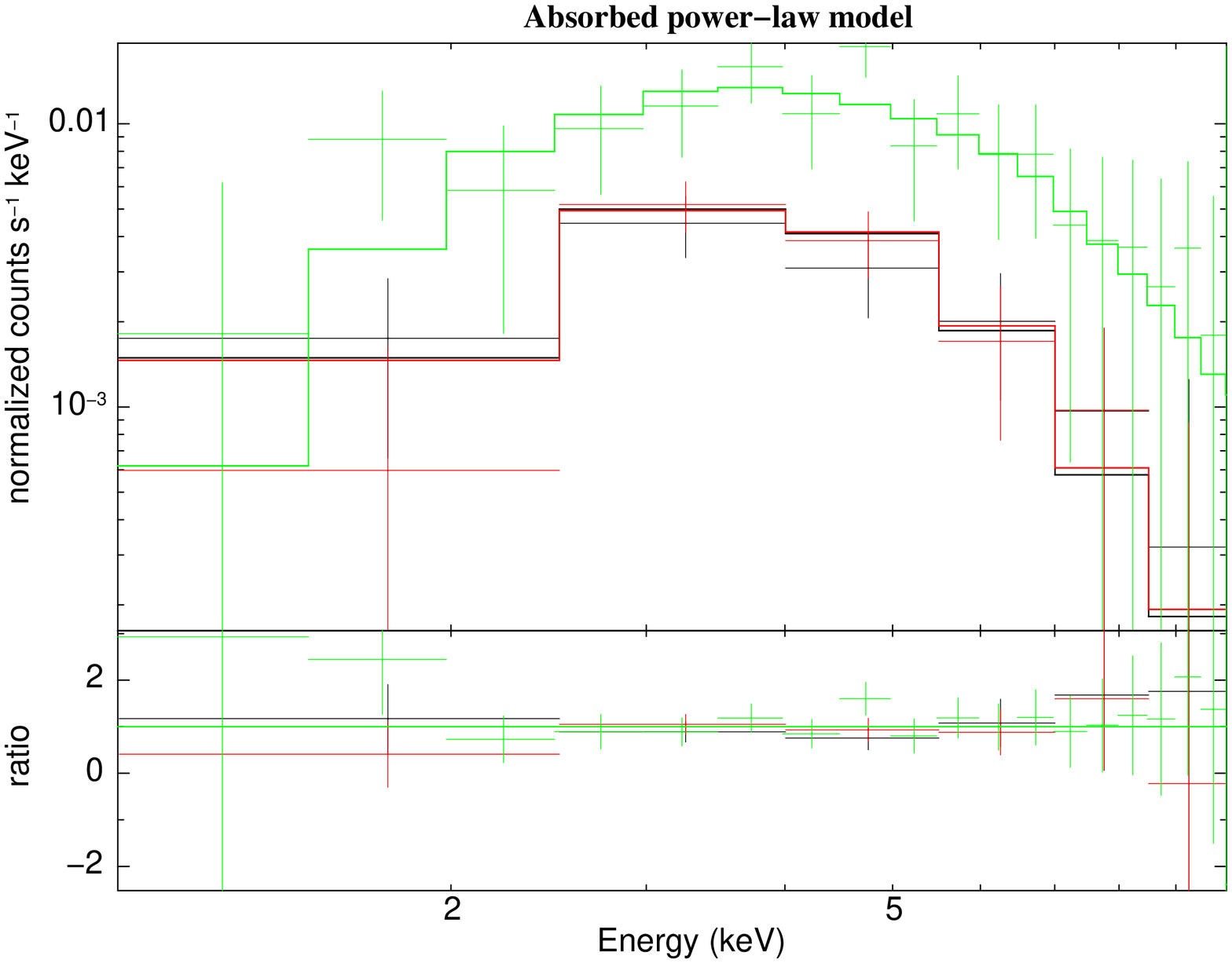}
  \includegraphics[width=0.6\textwidth, angle=270]{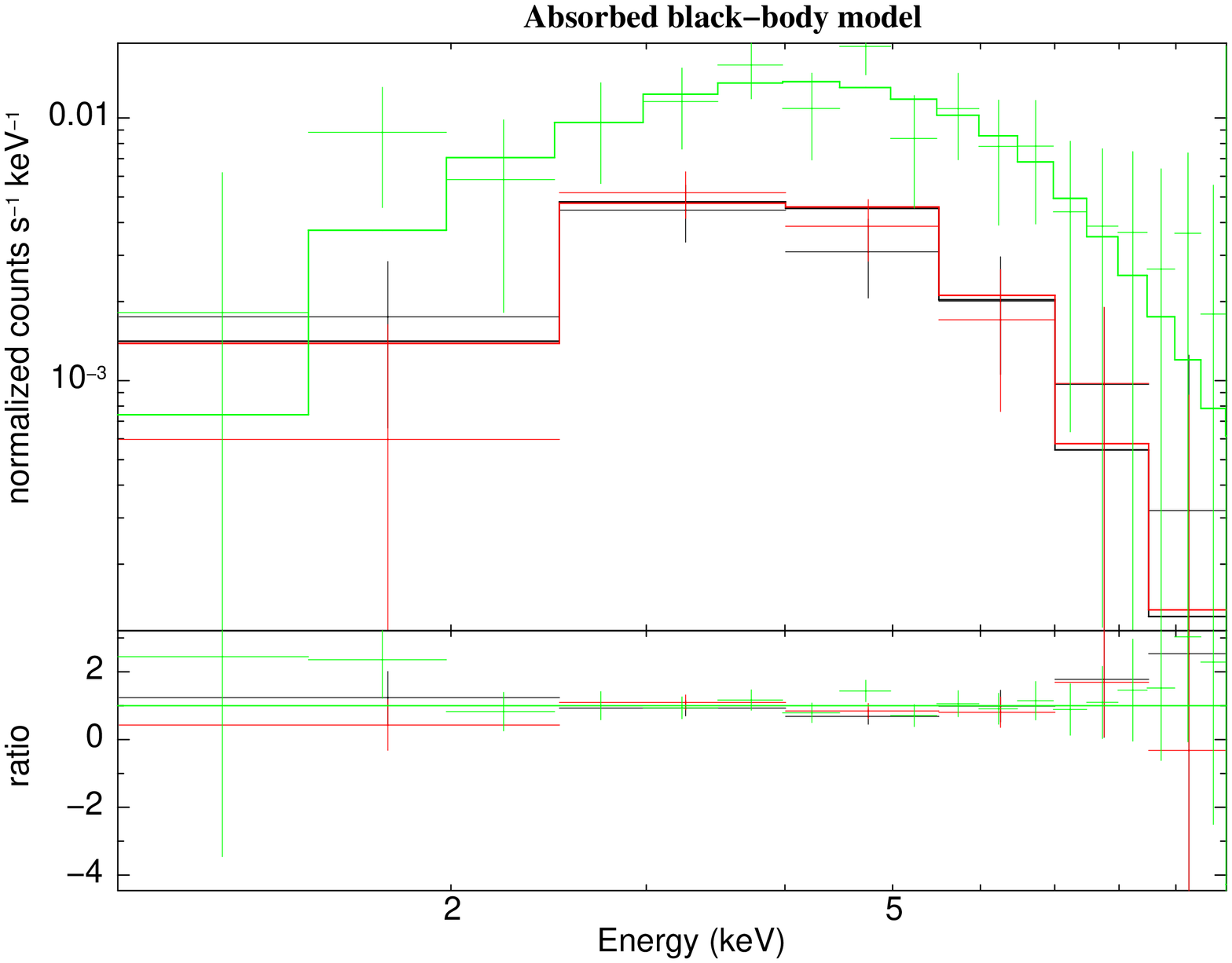}
  \caption{{\emph{XMM-Newton}} X-ray energy spectrum of \xmm\ for two
  different spectral models, an absorbed powerlaw ({\bf{top}}) and an
  absorbed black-body ({\bf{bottom}}). All detectors have been
  simultaneously fit (EMOS1: black, EMOS2: red, EPN: green) for an
  extraction radius of 75\arcsec. The plot shows flux points with a
  minimum significance of $5 \sigma$. The straight lines shows the
  combined fit to the data, folded with the instrument response
  function of the corresponding detector). The lower panel of each
  plot shows the residuals of the fit, illustrating the good match of
  both fit functions to the data.
    \label{fig::XMMSpectrum}}          
\end{figure}

The power-law fit yields a hard photon index of $1.74 \pm 0.1$, the
source is rather faint with a total detected (i.e. absorbed) flux in
the 2--10~keV range of $6.6 \times 10^{-13}$ erg cm$^{-2}$
s$^{-1}$. The column density $n_{\mathrm{H}} = 6.1 \times 10^{22}$
cm$^{-2}$ is marginally compatible with the average column density
through the Galaxy in this region ($n_{\mathrm{Gal H}} = 2.2 \times
10^{22} \mathrm{cm}^{-2}$ as reported by \citet{DickeyLockman}). The
black-body model yields a lower column density $n_{\mathrm{H}} = 3.6
\times 10^{22}$ cm$^{-2}$. It should however be noted, that the fit
parameters ($n_{\mathrm{H}}$ and $kT$) are somewhat correlated. Both
models fit the data well as shown in
Figure~\ref{fig::XMMSpectrum}. Comparing the values of the reduced
$\chi^2$, no spectral model can be preferred over the other. While the
results for the different background estimates are compatible within
statistical uncertainties, the scatter might give a hint on the
systematic error steming from the choise of the background and thus
the systematic error was estimated to 0.2 on the photon index and 20\%
on the integrated flux. The result of the spectral analysis of the
{\emph{XMM-Newton}} data presented here is in agreement with the ASCA
data, given the large error on the latter spectral results (photon
index of $3.0^{+1.1}_{-0.9}$, $F_{2-10{\mathrm{keV}}} = 1.2\times
10^{-12}$ erg cm$^{-2}$ s$^{-1}$, $n_{\mathrm{H}} = 9.6^{+4.7}_{-3.3}
\times 10^{22} \mathrm{cm}^{-2}$). The spectral analysis shows
however, that the photon index is rather hard compared to the ASCA
spectrum.

\section{Interpretation of the Multi-wavelength data}

The {\emph{XMM-Newton}} data reveal the source \xmm\ in positional
coincidence with the unidentified ASCA source \asca. The source
exhibits a rather hard power-law spectrum of photon index $\sim 1.75$,
although a black-body spectrum of the X-ray emission with a
temperature 1.56~keV, corresponding to $1.8 \times 10^{7}~K$ provides
an equally good fit. The source is relatively faint
($F_{2-10~\mathrm{keV}} = 6.6 \times 10^{-13}$ erg cm$^{-2}$ s$^{-1}$)
and shows a rather strong absorption of $n_{\mathrm{H}} = 6.1 \times
10^{22}$ cm$^{-2}$. \xmm\ is extended in nature with a compact core
and a faint tail, resembling in morphology and spectral properties
typical Pulsar Wind Nebulae (PWN). Asymmetric ``trails'' in PWN can be
generated either a) through an asymmetric density distribution of the
surrounding medium, preventing the expansion of the PWN on one side as
e.g.\ seen in Vela~X, b) dynamically by a supersonic motion of the
pulsar with respect to the ISM, generating a bow-shock and a
``cometary trail'' \citep[for a recent review of PWN
see][]{GaenslerReview}. The hypothesis that \xmm\ is a PWN (stemming
from the hard energy spectrum and the morphology) is strengthened
further by the location of this object within the boundaries of the
shell-type radio SNR \snr. This PWN scenario could finally be
confirmed by the detection of a pulsar coincident with \xmm. No
pulsation has been found in the current {\emph{XMM-Newton}} data and
no known radio pulsar is located towards \xmm\ in the current ATNF
catalogue~\citep{ATNF}. Future deep radio or X-ray timing observations
might serve to detect the associated pulsar. The ``beaming fraction'',
i.e.\ the faction of 4 $\pi$ steradians covered by the pulsar beam
during one rotation, corresponds to the probability that the beam
sweeps the line-of-sight of an observer. So far, no agreement has been
reached on the relation between the beaming fraction and the period of
the pulsar, and estimates range from beaming fractions of 30\% to
100\% for a typical 100~ms pulsar~\citep{Narayan, Lyne, Tauris}. In
any case, the potential pulsar to the PWN discussed here might be
beamed away from our line-of-sight. In this case, the confirmation of
the PWN picture must come from the detection of electron cooling,
resulting in a softening of the X-ray spectrum away from the pulsar as
seen in many other PWN \citep[e.g.\ in
\object{G\,21.5--0.9,}][]{SlaneG21.5}). Given the faintness of \xmm\
the current data do not allow for the detection of such an
effect. Very deep high angular-resolution {\emph{Chandra}} studies
will finally confirm or rule out this scenario. Nevertheless, the
{\emph{XMM-Newton}} data are very suggestive of a PWN, since the hard
X-ray spectrum, the extended nature, the position within an radio SNR
and the spatial coincidence with a VHE $\gamma$-ray source seem
unlikely to occur by chance. In the following discussion we will
therefore assume that the emission in the radio, X-rays and VHE
$\gamma$-rays are connected and characterise a {\emph{composite SNR}}.

\begin{figure}
  \centering
  \includegraphics[width=0.9\textwidth]{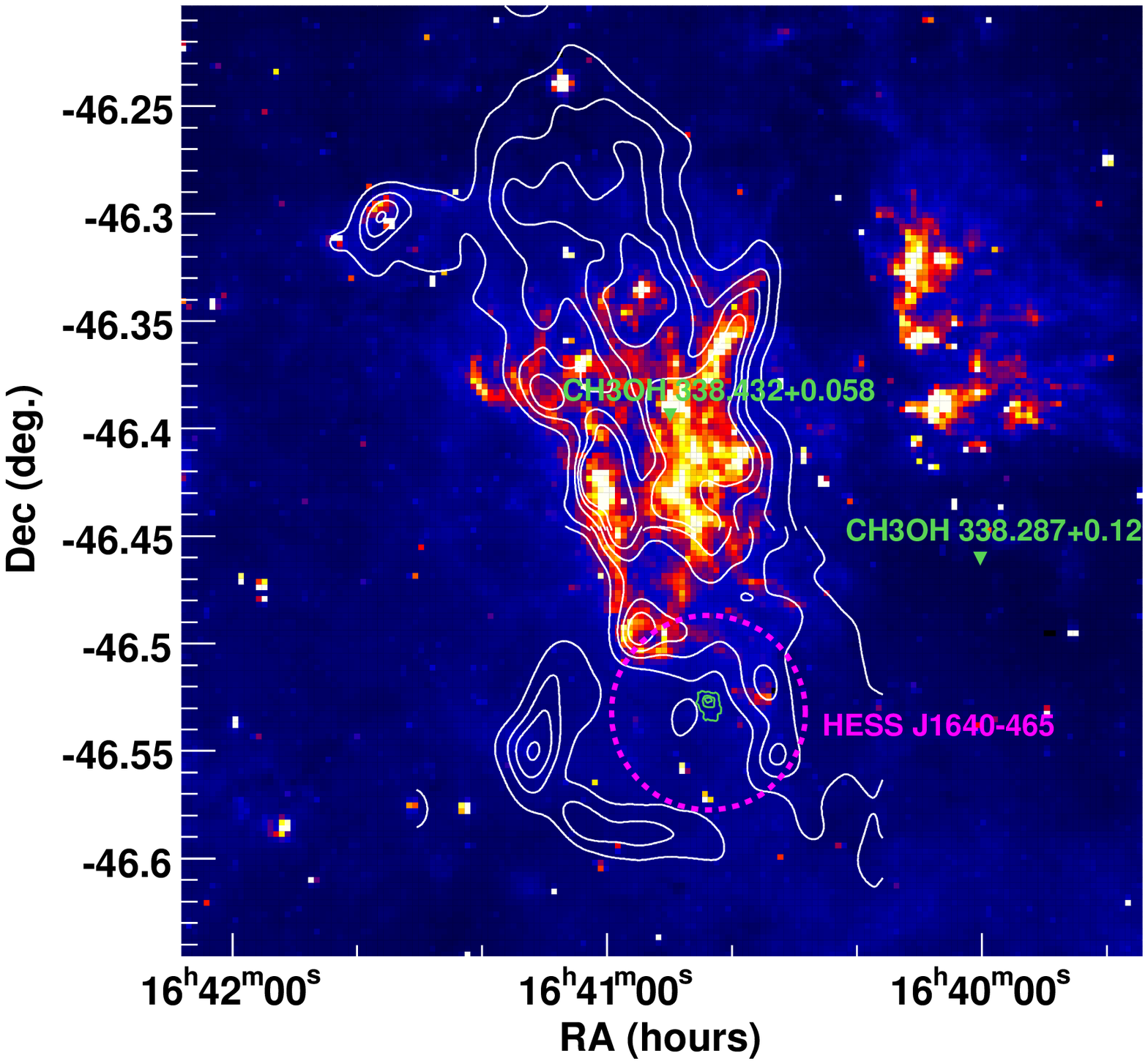}
  \caption{{\emph{Spitzer}} Space telescope data of the GLIMPSE survey
  at 8~$\mu$m~\citep{Spitzer} along with 843~MHz MOST radio data shown
  as white contours. As described in~\citet{Whiteoak}, the so-called
  radio-bridge extending to the north of \snr\ coincides very well
  with a bright HII-region in which dust emission shows ongoing star
  formation. This region is located at a distance of $\sim 3$~kpc and
  might have played an important role in the development of the SNR if
  both objects are located at the same distance. Also shown are the
  position and extension of \this\ (pink) and the combined MOS1 and
  MOS2 {\emph{XMM-Newton}}X-ray contours (green) all coinciding with
  the radio SNR \snr\ described in the text. Also shown are the known
  Masers in the region, indicating an interaction between shock waves
  and the ambient medium and illustrating that the region is still
  active in terms of star formation.
  \label{fig::Spitzer}}          
\end{figure}

The morphology of \this\ seems to bear remarkable similarity with that
of HESS\,J1813--178~\citep{HESSScan, HESSScanII, Funk1813}. In both
cases there is a coincidence of an extended TeV $\gamma$-ray source
with an extended hard spectrum X-ray source located within the radio
shell of an SNR. As in the case of HESS\,J1813--178 it is not clear,
whether the $\gamma$-ray emission originates in the shell of the SNR,
or from the core, since the angular resolution of VHE $\gamma$-ray
instruments is too coarse to resolve the small structures seen in
X-ray and radio observations. The ``Gaussian equivalent width'' of the
radio SNR can be defined as $\sigma_{\mathrm{SNR}} = (\sigma_x^2 +
\sigma_y^2 - r_{\mathrm{smooth}}^{2})^{1/2}$ ($\sigma_x$ and
$\sigma_y$ the Gaussian width in RA and Dec direction of the radio
emission region smoothed with a Gaussian of size
$r_{\mathrm{smooth}}=1.2\arcmin$ as used in the \HESS\
analysis). This ``equivalent width'' of the radio shell of \snr\ is
$2.5\arcmin \pm 0.2\arcmin$, perfectly compatible with the size of the
VHE $\gamma$-ray emitting region with an rms of $2.7\arcmin \pm
0.5\arcmin$. As in the case of HESS\,J1813--178 both the radio shell
and the central X-ray core must therefore be considered as viable VHE
$\gamma$-ray emitters. However, there are several notable differences
to HESS\,J1813--178:
\begin{itemize}
\item The X-ray emission of \this\ is much weaker. This has several
important consequences: any hard X-ray counterpart falls well below
the sensitivity limit of current detectors such as INTEGRAL and
Suzaku. Thus no hard X-ray counterpart to \xmm\ exists as seen for
HESS\,J1813--178 (suggesting acceleration to PeV energies). The faint
X-ray emission also does not allow for an exclusion of a
{\emph{thermal origin}} of the keV emission. Future deep X-ray
observations might finally distinguish between a power-law or a
black-body emission model. 
\item An unidentified EGRET source (\eg)~\citep{EGRETCat} is in
positional coincidence with \snr. If the sources are associated, this
increases the $\gamma$-ray power even further.
\item So far, no dedicated search for a radio pulsar has been
  performed within the remnant and thus so far no pulsar has been
  found in \snr. As discussed above, the pulsar light-cone might not
  sweep us as observers. There is however a weak radio point-source
  located at $16^{\mathrm{h}}40^{\mathrm{m}}47.75^{\mathrm{s}}$,
  $-46^{\mathrm{d}}32\arcmin03.3\arcsec$ within the remnant, that
  could possibly be identified as a pulsar or even a ``relic'' PWN in
  deep observations. This source, which can be seen in
  Figure~\ref{fig::XMMFoVZoom}, lies $\sim 40\arcsec$ from \xmm.
\item The distance to \this\ ($\sim 8.6$~kpc) is only poorly known,
  constrained only by the $\Sigma-D$ relation for \snr\ (with $\Sigma$
  being the surface brightness and $D$ the distance). Such estimates
  are notoriously unreliable~\citep{Green}. The nearby HII-regions
  shown in Figure~\ref{fig::Spitzer} have measured radial velocities
  and implied distances of 3--4 kpc~\citep{Whiteoak}, and \snr\ may be
  associated with this region of star formation. The relatively strong
  \HESS\ and EGRET fluxes (provided that these objects are associated)
  would also argue for a distance $\ll 10$ kpc on energetics
  grounds. Nevertheless, in the following a distance of $\sim 8$ kpc
  will be assumed.
\end{itemize}

\noindent Whilst several morphological properties of \this\ are very
similar to HESS\,J1813--178, spectral properties differ quite strongly
and the first two differences listed above are reminiscent of another
VHE $\gamma$-ray source: HESS\,J1825--137~\citep{HESSJ1825,
HESSJ1825II}. This object is the only known VHE $\gamma$-ray source to
exhibit energy dependent morphology~\citep{HESSJ1825II}. For
HESS\,J1825--137, an X-ray PWN was found in {\emph{XMM-Newton}}
observations of the energetic pulsar
PSR\,J1826--1334~\citep{Gaensler}, extending asymmetrically to the
south of the pulsar. Similarly, the VHE $\gamma$-ray emission shows an
asymmetric emission, extending to the south of the pulsar, however on
a completely different scale than the X-ray emission (the X-ray
emission extends 5\arcsec, whereas the $\gamma$-ray emission extends
$\sim 1^{\circ}$ to the south). Taking into account the different loss
timescales of the $\gamma$-ray and X-ray emitting electrons, the
situation can be plausibly explained by the $\gamma$-rays being
generated by lower energy electrons than the X-rays. This scenario was
strengthened by the detection of energy dependent morphology in the
VHE $\gamma$-ray source, leading to a convincing identification of
this object as the PWN of PSR\,J1826--1334~\citep{HESSJ1825II}.
PSR\,J1826--1334 is a powerful ``Vela-like'', $\sim$20 kyr old
pulsar. For such evolved pulsars, in which the spin-down luminosity
has changed substantially during the lifetime, braking effects are no
longer negligible and a decrease of the particle injection rate with
age must be taken into account.  According to~\citet{Pacini}, the
spin-down luminosity $\dot{E} (t)$ evolves in time as
\begin{equation}
\label{eq::evolution}
\dot{E}(t) = \dot{E} (0) \left( 1+\frac{t}{\tau_0} \right)^{-
\frac{n+1}{n-1}}, \qquad \mathrm{with} \quad \tau_0 =
\frac{P_0}{\dot{P}_0 (n-1)}
\end{equation} 

\noindent (with $n$ the braking index and $\tau_0$ the initial
spin-down timescale of the pulsar). If \this\ indeed represents a PWN
similar to HESS\,J1825--137, a relatively powerful, $10^{4}-10^{5}$
year old pulsar should exist within \snr. The age of \snr\ has been
estimated to $\sim 2 \times 10^{4}$ years, assuming expansion in the
Sedov phase into a medium of density $10$ cm$^{-3}$, and using the
distance estimate of $\sim 8$~kpc.

Taking into account the examples of both HESS\,J1813--178 and
HESS\,J1825--137, in the following the multi-frequency properties of
\this\ will be discussed with the aim to connect the radio, X-ray and
VHE $\gamma$-ray emission into a consistent
picture. Table~\ref{tab::XraysGammarays} summarises the properties of
the X-ray and of the $\gamma$-ray emission.

\begin{table*}[h]
    \caption{Comparison between size and spectral properties of \xmm\
    and \this\ for an assumed distance of 8~kpc. \vspace{0.4cm}}
    \label{tab::XraysGammarays}
    \centering
    \begin{tabular}{ c | c  c }
      \tableline
      & \xmm & \this \\ 
      \tableline
      Angular Size &  $0.45\arcmin \pm 0.05\arcmin$ & $2.7\arcmin \pm 0.5\arcmin$\\
      Linear Size (pc)& $\sim 1 d_{8 \mathrm{kpc}}$ & $\sim 6 d_{8 \mathrm{kpc}}$\\
      Apparent Luminosity ($\times 10^{33}$ erg/s) & $5 d_{8
      \mathrm{kpc}}^2$& $170 d_{8 \mathrm{kpc}}^2$\\ 
      Photon Index & $1.75 \pm 0.12_{\mathrm{stat}}$ & $2.42 \pm
      0.15_{\mathrm{stat}}$ \\ 
      \tableline
    \end{tabular}
\end{table*}

\begin{figure}
  \centering
  \includegraphics[width=0.9\textwidth]{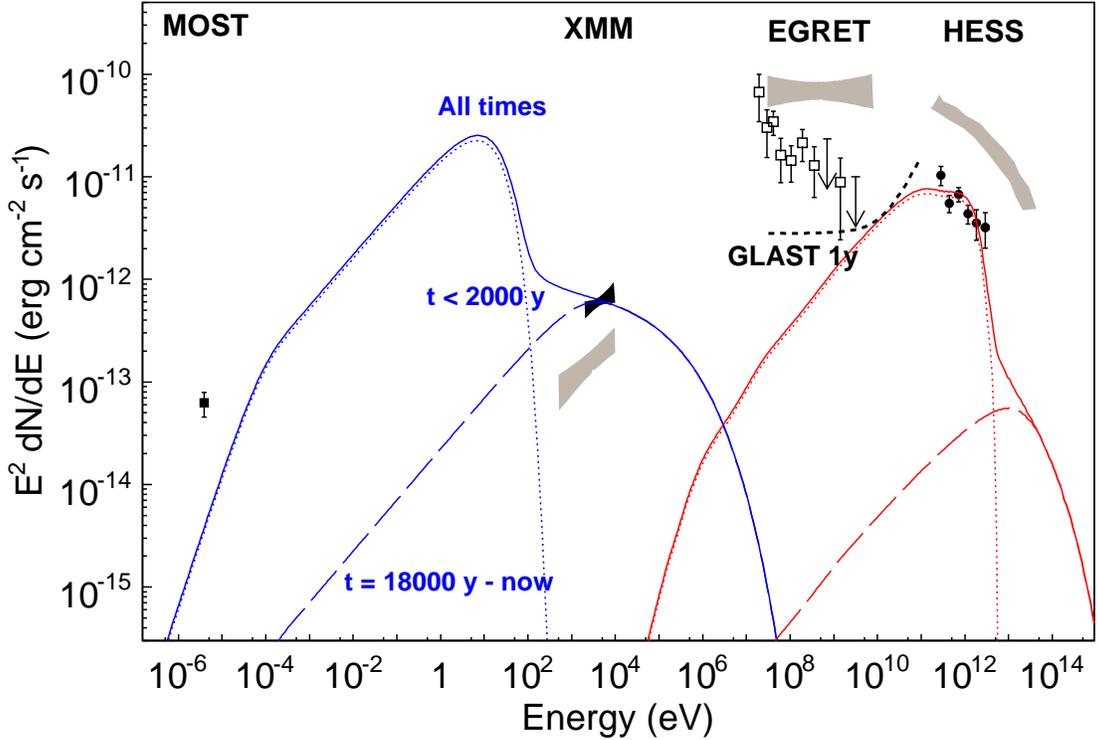}
  \caption{Spectral energy distribution of \this\ for the model with a
    time dependent rate of injection of relativistic electrons as
    described in the text. For comparison the spectral data for the
    VHE $\gamma$-ray PWN HESS\,J1825--178 are shown as grey bands. The
    {\emph{XMM-Newton}} data between 2 and 10~keV are corrected for
    the absorption and a systematic error band of 0.2 on the spectral
    index and 20\% on the flux level has been added. The different
    curves indicate emission from electron populations injected at
    different time slices in the age of the pulsar (assumed to be
    20,000 years of age). The injection rate was assumed to vary in
    accordance with the change in spin-down power as described in the
    text. Youngest electrons (injected in the last 2000~years)
    responsible for the X-ray emission are shown as dashed lines,
    oldest electrons (injected in the first 2000~years after the
    pulsar's birth) are shown as dotted lines, dominating the VHE
    $\gamma$-ray emission. The red curves show the Inverse Compton
    component, whereas the blue lines show the synchrotron emission in
    the chosen model. The radiation fields for the inverse Compton
    emission are assumed to be nominal Galactic radiation fields taken
    from~\citet{PorterMoskalenko}. The model parameters are specified
    in the text. The black dashed line shows the 1-year sensitivity
    curve for the GLAST-LAT taking into account the diffuse Galactic
    and the residual instrumental background.
  \label{fig::SED}}    
\end{figure}

Figure~\ref{fig::SED} compares the spectral energy distributions of
\this\ and HESS\,J1825--137. The main characteristics that have to be
explained by the modelling are that the VHE $\gamma$-ray source is an
order of magnitude larger than the X-ray source and that the VHE
$\gamma$-ray power is more than an order of magnitude larger than the
X-ray power. As in the case of HESS\,J1825--137, it is rather
difficult to establish a multi-wavelength connection in a leptonic
one-zone model~\citep{AhaAto99} with constant injection and a single
population of electrons responsible for the whole emission.  For
plausible radiation and magnetic field energy densities, such models
generally predict X-ray synchrotron fluxes equal to or greater than
the associated TeV IC fluxes, unless suppressed magnetic fields or
rather peculiar environments like in the Galactic Centre are
invoked~\citep{HintonAha}. As it seems hardly possible to find
satisfactory parameters for such a simple model, a
\emph{time-dependent} rate of injection of electrons into the nebula
will be considered. For the following argument, the existence of a
$\sim 2 \times 10^4$ year old pulsar produced in the supernova
explosion of \snr\ will be assumed. For younger ages, no satisfactory
fit to the data can be found in the frame of the model described
here. It will be further be assumed that the size of the X-ray source
is limited by electron synchrotron cooling which in turn limits the
age of particles seen at X-ray energies. To explain the different
sizes of the emission regions in X-ray and VHE $\gamma$-rays, cooling
times of the X-ray emitting electrons must be short in comparison to
the age of the source, while cooling times for the VHE $\gamma$-ray
emitting electrons must be long in comparison to the age of the
source.  This situation is rather likely for a system of this age: For
a typical $10~\mu$G field, 4~keV synchrotron X-rays are generated by
$\sim 100$~TeV electrons, which lose their energy on timescales of
$\sim$1,200~years. In contrast, 0.8~TeV IC $\gamma$-ray are emitted by
$\sim 3$~TeV electrons, which cool on timescales of $\sim 32,000$
years. A natural explanation for the different sizes of the emission
regions is to invoke different populations of electrons with different
cooling timescales and hence propagation distances. In this scenario
the size of the X-ray nebula is limited by electron cooling on $\sim
1000$ year timescales, in contrast the angular size of the
$\gamma$-ray source reflects the propagation speed of electrons
injected soon after the birth of the pulsar $\sim 2 \times 10^{4}$
years ago.

The curves in Figure~\ref{fig::SED} illustrate the model scenario
invoked to connect \xmm\ to \this\ for populations of electrons
injected at different slices in time $t$ during the lifetime of the
pulsar. A canonical $E^{-2}$ injection spectrum was used with a lower
energy cut-off at 10~GeV and an exponential cut-off at 1~PeV, as the
rather hard X-ray spectrum yields a lower limit for the cutoff energy
of a few hundreds of TeV. A time-independent magnetic field strength
of 10 $\mu$G was assumed for the whole nebula. The injection rates of
electrons varies with time, assuming that the power injected into
electrons follows the spin-down power of the pulsar which varies
according to Equation~\ref{eq::evolution} (with $n=3$ and $\tau_0 =
300$ years). The dotted line shows the electrons injected in the first
2000 years after the birth of the pulsar (20,000 years ago),
i.e. ``old'' electrons. The injected electron population evolves in
time, taking into account synchrotron and IC energy losses. The
population of ``old'' electrons totally dominates the synchrotron
emission below X-ray energies and is responsible for the VHE
$\gamma$-rays detected in \this. The populations of ``youngest''
electrons (dashed curve) injected in the last 2000 years are less
drastically cooled, as can be seen in Figure~\ref{fig::SED}.
According to the model shown here, these youngest electrons are
responsible for the bulk of the X-ray emission as seen in the compact
X-ray source.  To explain the X-ray source as synchrotron emission
from a PWN, the required (present day) luminosity in accelerated
electrons is $1.4 \times 10^{36} (\mathrm{B}/10\mu\mathrm{G})^{-2}
(\mathrm{d}/8\mathrm{kpc})^2$ erg/s, or about 50\% of the spin-down
power of a pulsar like PSR\,J1826--1334 ($3 \times 10^{36}$ erg/s),
assumed to be a representative of the pulsars powering these VHE
$\gamma$-ray PWN systems. The total electron luminosity integrated
over the lifetime of the pulsar amounts to $1.4 \times 10^{48}$
ergs. An upper limit from the 8~$\mu$m GLIMPSE {\emph{Spitzer}} data,
integrating within the area of the VHE $\gamma$-ray source, yields an
unconstraining upper limit of $\sim 1 \times 10^{-8}$ erg cm$^{-2}$
s$^{-1}$ (at $\sim 0.2$ eV). It should however be noted that in the
derivation of this limit no subtraction of prominent IR sources has
been applied. Although the figure shows how to accommodate the
different sizes and fluxes of the X-ray and VHE $\gamma$-ray
production regions, it must still be seen as an oversimplified
picture. A realistic model requires the inclusion of the time (and
space) dependence of the magnetic field, as well as a proper treatment
of the propagation of particles within the nebula.

Also apparent from these model curves is that it is rather difficult
to connect \this\ and \xmm\ with the unidentified EGRET source \eg\
within the framework of an IC scenario. However, an association of
\eg\ and \this\ is rather natural in two alternative scenarios: a
model in which the VHE $\gamma$-rays are generated by electron
Bremsstrahlung or in a hadronic emission model. If the ISM density in
this region is high ($n$ is more than about 100 cm$^{-3}$), then
Bremsstrahlung may be the dominant $\gamma$-ray emission mechanism at
TeV energies for a population of ultra-relativistic electrons. If
\this\ and \eg\ are indeed associated, Bremsstrahlung may be the most
natural mechanism to explain the high energy part of the SED.

Alternatively, the EGRET and \HESS\ detected $\gamma$-rays could be
generated by hadronic interactions of accelerated protons via
$\pi^0$-decays. In this case the 2--10~keV X-ray flux can be connected
to these accelerated hadrons by secondary electrons generated in the
hadronic interactions. Assuming a distance of $8$~kpc, the measured
$\gamma$-ray flux above 200~GeV suggests a total energy in the
accelerated hadrons of $W_p = 7.6 \times 10^{50} d/(8~\mathrm{kpc})
n/(1 \mathrm{cm}^{-3})$ erg. This energy represents a plausible
injection from a single SNR for values of $n$ more than 1~cm$^{-3}$
and/or distances significantly closer than 8~kpc.  However, in this
scenario, secondary electrons would be produced in the same region as
the TeV $\gamma$-rays, and similar \emph{sizes} for the X-ray and
$\gamma$-ray sources might naively be expected. The observed factor of
six difference in size therefore counts against an association of
\xmm\ and \this\ in this scenario. Whilst a hadronic scenario for
$\gamma$-ray emission remains viable, if \xmm\ and \this\ are
unrelated, then the excellent positional agreement between these two
sources can be considered somewhat surprising.

Given the fact that there are several other prominent candidates for
\eg\ (as described above) and that the EGRET source is the least well
measured object in this context, it seems that a strong claim of a
possible connection has to await the launch of the upcoming GLAST
satellite. The dashed line in Figure~\ref{fig::SED} shows the 1-year
sensitivity of the GLAST-LAT, including the diffuse Galactic
background and the instrumental background. The curve demonstrates
that the LAT is well suited to shed new light on this region in the
MeV to GeV range via its superior angular resolution (and sensitivity)
in comparison to EGRET.

\section{Summary and conclusion}

The detailed {\emph{XMM-Newton}} X-ray data taken towards \this\ show
a hard-spectrum extended object towards the centre of the SNR \snr,
suggesting the detection of a new composite SNR in which the X-rays
are generated by synchrotron emission from a central PWN. The ultimate
proof of this scenario requires the detection of a coincident pulsar,
in either X-ray or radio observations, or the detection of spectral
cooling characteristic of PWN systems. Both the shell of the SNR and a
central PWN are viable VHE $\gamma$-ray emitters and the two scenarios
cannot be distinguished at this point (the situation is similar to the
case of HESS\,J1813--178). If the VHE $\gamma$-ray are generated by
inverse Compton emission associated with a PWN, the spectral energy
distribution shows an interesting similarity to another VHE
$\gamma$-ray PWN HESS\,J1825--137. As is the case for that object the
low ratio of the X-ray power to the VHE $\gamma$-ray power in \this\
suggests a time dependent rate of injection for the relativistic
electrons responsible for the X-ray emission and an older (and more
numerous) population of electrons producing the VHE $\gamma$-ray
emission. IR to soft X-ray measurements of the synchrotron nebula
associated with \this\ could confirm this picture. However, any
synchrotron emission from \this\ in this energy range appears to be
buried beneath thermal emission. An extension of the $\gamma$-ray
spectrum to higher energies (for example from a deeper observation
with \HESS) would allow us to probe the IC emission of the electrons
responsible for the $> 1$ keV synchrotron emission.  The detection of
the extended X-ray source in the centre of the Supernova remnant \snr\
{\emph{following}} the detection of a VHE $\gamma$-ray source
demonstrates that X-ray follow-up observations of TeV $\gamma$-ray are
well motivated and provide important insights into the nature of these
objects.

\acknowledgments The authors would like to acknowledge the support of
  their host institutions, and additionally support from the German
  Ministry for Education and Research (BMBF). SF acknowledges support
  of the Department of Energy (DOE), JAH is supported by a PPARC
  Advanced Fellowship. SW and GP acknowledge support from BMBF through
  DESY and DLR grants. We would like to thank the whole \HESS\
  collaboration for their support. We would also like to thank Teddy
  Cheung for the reanalysis of the MOST radio data and helpful
  discussions on this source.

\clearpage

\end{document}